\documentclass[12pt]{iopart}

\usepackage{amssymb}
\usepackage[dvips]{graphicx}
\usepackage{latexsym}
\usepackage{amssymb}
\usepackage{cases}
\usepackage[compress]{cite}
\usepackage{xcolor}
\usepackage{enumerate}
\usepackage{mathrsfs}

\newtheorem{lemm}{Lemma}

\begin{document}
\title[Effects of nodes position on diffusion and trapping efficiency in FSF trees]{Effects of node position on diffusion and trapping efficiency for random walks on fractal scale-free trees}

\author{Junhao Peng $^{1,2}$ , Guoai Xu$^3$}
\address{1.College of Math and Information Science , Guangzhou
University , Guangzhou 510006 , People¡¯s Republic of China. \\
2.Key Laboratory of Mathematics and Interdisciplinary Sciences of Guangdong
Higher Education Institutes, Guangzhou University,Guangzhou 510006 ,China.
3.State Key Laboratory of Networking and Switching Technology, Beijing University of Posts and Telecommunications , Beijing 100876 , People¡¯s Republic of China.
} \ead{\mailto{pengjh@gzhu.edu.cn}}

\begin{abstract}
We study unbiased discrete random walks  on the FSFT based on the its self-similar structure  and the relations between  random walks and electrical networks. First, we  provide new methods to derive analytic  solutions of the MFPT for any pair of nodes, the MTT for any target node and MDT  for any starting node.
 And then, using the MTT and the MDT as the measures of trapping efficiency and diffusion efficiency respectively,  we analyze the effect of trap's position on trapping efficiency and the effect of starting position on diffusion efficiency. Comparing the trapping efficiency and diffusion efficiency among all nodes  of FSFT, we find the best (or worst)  trapping sites  and the best (or worst) diffusing sites. Our results show that: the node which is at the center  of FSFT is the best trapping site, but it is also the worst diffusing site. The   nodes which are the farthest nodes from the two hubs are the worst trapping sites, but they are also the best diffusion sites.  Comparing the maximum  and minimum of MTT and MDT, we found that the maximum  of MTT is almost $\frac{20m^2+32m+12}{4m^2+4m+1}$ times of the minimum of MTT, but the   the maximum  of MDT is almost equal  to the minimum  of MDT. These results shows that the position of target node has big effect on trapping efficiency, but the position of starting node almost has no effect on diffusion efficiency. We also conducted numerical simulation to test the results we have  derived, the results we derived are consistent with those obtained by numerical simulation.
\end{abstract}

\maketitle
\section{Introduction}
  Classic fractals exhibits many properties of reality systems  such as  scale free\cite{BaAl99, JuKiKa02} and  small-world properities\cite{Watts98,  ZhangZhou07}. It is good model to mimic  reality systems\cite{SongHaMa05, SongHaMa06, RoHa07}. Random walks on fractals , which can be applied as model for transport in disordered media\cite{HaBe87, Avraham_Havlin04},  has attracted   lots of interests\cite{VoRe10, MariSaSt93, RaTo83, BeTuKo10}.  The range of applicability and of physical interest is enormous \cite{LlMa01, We84, JiYang11,  Ko00, ZhZh08}.\par
A basic quantity relevant to random walks is the trapping time or mean first-passage time (MFPT), which is the expected number of steps to hit the target node(or trap) for the first time, for a walker starting from a source node. It is a quantitative indicator to characterize the transport efficiency and  many other quantities can be expressed in terms of it. Locating the target node(or trap) at one special node and average the MFPTs over all the starting nodes, we get mean trapping time(MTT) for the special node. Locating the source node at one special node and the average the MFPTs over all the target nodes,  we obtain mean diffusing time(MDT) for the special node. Both the MTT and MDT  varies with the position of node and they can be used as the measures of trapping efficiency and diffusion efficiency for network nodes respectively. Comparing the  MTT and  MDT  among  all the network nodes, we can find the effects of node position on the trapping efficiency and diffusion efficiency. The nodes which  have the minimum MTT (or the maximum MTT) are best (or worst)  trapping sites and  the nodes which have the minimum MDT (or maximum MDT) are the best (or worst) diffusion sites .\par
  In the past several years, MFPT for random walks on   fractals  have been extensively studied\cite{BeTuKo10,  GiMaNa94, KoBa02, ShMaNa09,  MoHa89, BeMeTe08, HaRo08}. For example, the MTT for some special nodes were  obtained for different  fractals(or networks) such as Sierpinski  gaskets\cite{KoBa02}, Apollonian network\cite{ZhGuXi09}, pseudofractal scale-free web \cite{ZhQiZh09}, deterministic scale-free graph\cite{AgBuMa10} and some special trees\cite{CoMi10,  ZhZhGa10, LiZh13, LiWuZh11,  Agl08, ZhLiLin11}. The MDT for some special nodes  were obtained for exponential treelike networks\cite{ZhLiLin11}, scale-free Koch networks\cite{ZhGa11} and deterministic scale-free graph\cite{AgBu09}. There were also some works focusing on global mean first-passage time (GMFPT), i.e., the average of MFPTs over all pairs of nodes, these results were  obtain for some special trees \cite{ ZhZhGa10, LiZh13, LiWuZh11, ZhLiLin11, ZhWu10,  ZhYu09} and dual Sierpinski gaskets\cite{WuZh11}. \par
 However, the results of MTT and MDT which were obtained are only restricted to some special nodes and we can not compare the trapping efficiency and diffusion efficiency among all the network nodes.  It is still difficult to deriving the analytic solutions of  the MTT for any target node(or trap) and the MDT for any source node in these networks.   \par
As for the recursive fractal scale-free trees(FSFT),  the MTT for the hub node and the GMFPT had been obtained\cite {ZhLi11}. The MTT for some low-generation nodes can also be derived due to the methods of Ref. \cite{MeAgBeVo12}. But the  analytic calculations of MFPT for any pair of nodes, the MTT for any target node and the MDT for any source node were still unresolved.
\par
 In this paper,  based on the self-similar structure of FSFT and the relations between  random walks and electrical networks\cite{Te91, LO93},  we  provide new methods to derive analytic  solutions of the MFPT for any pair of nodes, the MTT for any target node and MDT  for any starting node.\par
 Further more, using the MTT and the MDT as the measures of trapping efficiency and diffusion efficiency respectively,  we compare the trapping efficiency and diffusion efficiency among all nodes  of FSFT and find the best (or worst)  trapping sites  and the best (or worst) diffusing sites. Our results show that: the central node  of FSFT is the best trapping site, but it is also the worst diffusing site, the   nodes which are the farthest nodes from the two hubs are the worst trapping sites, but they are also the best diffusion sites.  Comparing the maximum  and minimum of MTT and MDT, we found that the maximum  of MTT is almost $\frac{20m^2+32m+12}{4m^2+4m+1}$ times of the minimum of MTT, but the   maximum  of MDT is almost equal  to the minimum  of MDT. These results shows that the position of target node has big effect on trapping efficiency, but the position of source node almost has no effect on diffusion efficiency.
  \section{Brief introduction to the FSFT }
\label{sec:1}
\begin{figure}
\begin{center}
\includegraphics[scale=0.3]{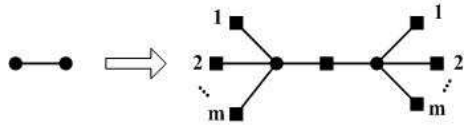}
\caption{Iterative construction method of the FSFT. For each edge of $G(t-1)$, we replace it by a cluster on the
right-hand side of the arrow, where  solid square stands for the  new nodes, while solid circle represents
the original node.}
\label{Edge_replace}       
\end{center}
\end{figure}

The recursive fractal scale-free trees(FSFT) we considered can be constructed iteratively\cite{JuKiKa02}. For convenience, we call the  times of iterations as the generation of the FSFT and  denote by $G(t)$ the FSFT of generation $t$. For $t = 0, G(0)$ is an edge connecting two nodes. For $t >0, G (t)$ is obtained from  $G(t-1)$ by performing the following operations on every edge as shown in Figure \ref{Edge_replace}: replace the edge by a path of 2 links long, with the two endpoints of the path being the same endpoints of the original edge and the new node having an initial degree 2 being  in the middle of the path, then attach m new nodes with the initial degree  1  to each endpoint of the path. \par 
\begin{figure}
\begin{center}
\includegraphics[scale=0.4]{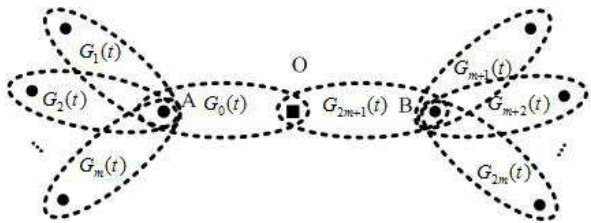}
\caption{Alternative construction of  FSFT which highlights self-similarity: the FSFT of generation $t$, denoted by $G(t)$, is composed of $2m+2$ copies of $G(t-1)$ which are labeled as  $G_0(t)$,  $G_1(t)$, $G_2(t)$, $\cdots$ ,$G_{2m+1}(t)$, and connected to one another at its two hubs $A$ and $B$.
}
\label{structure}
\end{center}
\end{figure}

 The  FSFT $G(t)$ can also be constructed by another method  highlighting self-similarity which is shown in Figure \ref{structure}\cite {ZhLi11}: the FSFT $G(t)$  is composed of $2m+2$ copies, called subunits, of $G(t-1)$ which are connected to one another by its two hubs (i.e., nodes with the highest degree).
\par
  This type of networks presents the following interesting structural features. They are scale free\cite{BaAl99, JuKiKa02} and  fractal with the fractal dimension $d_w= ln(2m + 2)/ln 2$\cite{SongHaMa05, SongHaMa06, RoHa07}, but they have not  small-world properities\cite{Watts98, RoHa07, ZhangZhou07}. According to its construction, one can easy obtain the total number of edges  $E_t$ and the total number of nodes  $N_t$ \cite{JuKiKa02, ZhLi11}.
 \begin{equation}
 E_t=(2m+2)^t
 \label{eq_edges}
 \end{equation}
\begin{equation}
N_t=1+E_t=1+(2m+2)^t
\label{eq_nodes}
\end{equation}
\section{Formulation of the problem}
\label{sec:gen_meth}
In this paper, we study  discrete-time random walks on FSFT $G(t)$. At each  step, the walker moves from its current location to any of its nearest neighbors with equal probability.  The quantity we are interested in is  mean first-passage time (MFPT), which is the expected number of steps to hit the target node(or trap) for the first time, for a walker starting from a source node.\par
Let $F(x,y)$  denote the MFPT from nodes $x$ to $y$ in FSFT $G(t)$ and $\Omega$ denote the node set of $G(t)$, the sum
$$k(x,y)=F(x,y)+F(x,y)$$ is called the commute time and the MFPT can be expressed in term  of commute times \cite{Te91}.
\begin{equation}
F(x,y)=\frac{1}{2}\left(k(x,y)+\sum_{u\in \Omega}\pi(u)[k(y,u)-k(x,u)] \right)
\label{FXY}
\end{equation}
where  $\pi(u)=\frac{d_u}{2E_t}$ is the stationary distribution for random walks on the   FSFT and $d_u$ is the degree of node $u$.\par
If we view the networks under consideration as electrical networks  by considering each edge to be a unit resistor and let $\Psi_{xy}$ denote the effective resistance  between two nodes $x$ and $y$ in the electrical networks,
 we have\cite{Te91}
\begin{equation}
k(x,y)=2E_t\Psi_{xy}
\label{KR}
\end{equation}\par
 where $E_t$ is the total numbers of edges of $G(t)$. Since the FSFT we studied  are trees, the effective resistance between any two nodes is exactly the  shortest-path length  between the two nodes. Hence
 \begin{equation}
 \Psi_{xy}=L_{xy}
 \end{equation}
 where $L_{xy}$ denote  the shortest path length between node $x$ to node $y$.  Thus
 \begin{equation}
k(x,y)=2E_tL_{xy}
\label{KL}
\end{equation}
 Substituting $k(x,y)$ with Eq.(\ref{KL})  in Eq.(\ref{FXY}), we obtain
\begin{eqnarray}
F(x,y)&=&E_t\left(L_{xy}+\sum_{u\in \Omega}\pi(u)L_{yu}-\sum_{u\in \Omega}\pi(u)L_{xu} \right)
\label{FXYL}
\end{eqnarray}\par
If we average the MFPTs over all the starting nodes and all target nodes, we  obtain MTT and MDT. That is to say, if we   define
\begin{eqnarray}
T_y&=&\frac{1}{E_t}\sum_{x\in \Omega,x\neq y}F(x,y) \label{MTTo}\\
D_x&=&\frac{1}{E_t}\sum_{y\in \Omega,y\neq x}F(x,y)  \label{MSTo}
\end{eqnarray}
$T_y$ is just the mean trapping time(MTT) for target node $y$ and $D_x$ is just mean diffusing time(MDT) for starting node $x$.
Let
\begin{equation}
S_x= \sum_{y \in \Omega}{L_{xy}}
\label{SX}
\end{equation}
\begin{equation}
W_x=\sum_{u\in \Omega}\pi(u)L_{xu}=\frac{1}{2E_t}\cdot \sum_{u \in \Omega}{(L_{xu}\cdot d_u)}
\label{WY}
\end{equation}
\begin{equation}
\Sigma=\sum_{u\in \Omega}\left(\pi(u)\sum_{x\in \Omega}L_{xu}\right)
\label{WS}
\end{equation}
 Substituting  $F(x,y)$ with Eq.(\ref{FXYL}) in Eqs.(\ref{MTTo}) and (\ref{MSTo}), we obtain
\begin{eqnarray} \label{MTT}
T_y&=&\sum_{x\in \Omega,x\neq y}\left(L_{xy}+\sum_{u\in \Omega}\pi(u)L_{yu}-\sum_{u\in \Omega}\pi(u)L_{xu} \right)  \nonumber  \\
&=&\sum_{x\in \Omega,x\neq y}L_{xy}+\sum_{x\in \Omega,x\neq y}\sum_{u\in \Omega}\pi(u)L_{yu}-\sum_{x\in \Omega,x\neq y}\sum_{u\in \Omega}\pi(u)L_{xu} \nonumber  \\
&=&S_y+N_t\cdot W_y-\Sigma
\end{eqnarray}
\begin{eqnarray} \label{MDT}
D_x
   &=&S_x+\Sigma-N_t\cdot W_x
\end{eqnarray}
Hence, if we can calculate $\Sigma$ and $S_x, W_x$ for any node $x$, we can calculate $F(x,y)$ for any two nodes $(x,y)$ and  MTT and MDT for any  node $x$. Although it is difficult to calculate these quantities for general tree, we presented methods for calculating these quantities for FSFT based on its self-similar structure. Therefore, we can calculate  MTT and MDT for any  node.
\section{The methods for calculating MTT and MDT}
\label{sec:det_meth}
 \subsection{Method for calculating $S_x$ and $W_x$}
 \label{Met_SW}
\begin{figure}
\begin{center}
\includegraphics[scale=0.7]{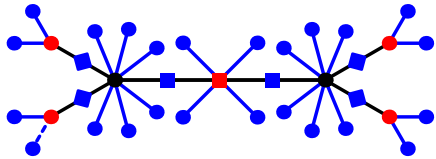}
\caption{The construction of the FSFT of generation $2$ for the limiting case of m = 2. The level information of its nodes: nodes colored black, level $0, 1, 2$; nodes colored red, level $1, 2$; nodes colored blue, level $2$. The subunit  represented by blue dotted line,   is labeled by a sequence $\{2, 4 \}$}.
\label{level_nodes}
\end{center}
\end{figure}

 For convenience, we classify the nodes of $G(t)$ into different  levels. Nodes, which are generated before $k$(include $k$)  times of iterations, are said to belong to level $k$ in this paper. Thus nodes which belong to level $k$ also belong to level $k+1$, $k+2$, $\cdots, t$. For example,  in the second generation FSFT with $m=2$, which  is shown in  Figure \ref{level_nodes}, the levels information of its nodes are: nodes  colored black belong to level $0, 1, 2$; nodes colored red  belong to  level $1, 2$; nodes colored blue belong to level $2$.\par
As shown in in Figure \ref{structure},  the FSFT $G(t)$  is composed of $2m+2$ subunits which are copies  of $G(t-1)$ and $G(t-1)$ is also composed of $2m+2$ subunits which are copies  of $G(t-2)$ . In order to tell apart the different  structures of these subunits, we  classify these subunits  into different levels and let $\Lambda_k$ denote the subunit of level $k(k\geq0)$. In this paper, $G(t)$ is said to be subunit of level $0$. For any   $k\geq0$, the $2m+2$  subunits of $\Lambda_k$ are said to be  subunits of level $k+1$. Thus, any edge of $G(t)$  is a subunit of level $t$ and $\Lambda_k$ is a copy of  FSFT with generation $t-k$.\par
\begin{figure}
\begin{center}
\includegraphics[scale=0.5]{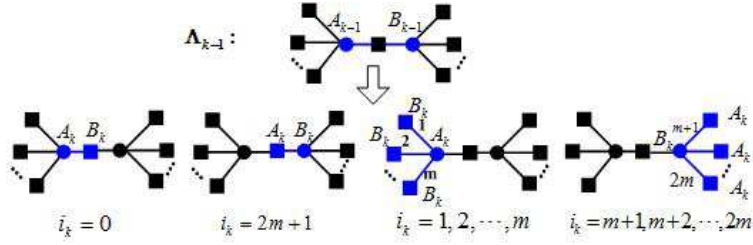}
\caption{Construction of $\Lambda_{k-1}$ and the relation  between the value of $i_k$ and the location of  subunit $\Lambda_{k}$ in $\Lambda_{k-1}$: subunit represented by blue line are the  subunits $\Lambda_{k}$ corresponding to value of $i_k$ below, whose two hubs are labeled as $A_k, B_k$. 
}
\label{lael_node}
\end{center}
\end{figure}

 In order to distinguish the subunits of different locations, similar to the method of Ref\cite{MeAgBeVo12}, we label the subunit  $\Lambda_k(1\leq k \leq t)$ by a sequence $\{i_1, i_2, ..., i_{k} \}$, where $i_j((1\leq j \leq k))$ labels its position in its father subunit $\Lambda_{j-1}$. Figure \ref{lael_node} shows the construction of $\Lambda_{k-1}$ and the relation  between the value of $i_k$ and the location of  subunit $\Lambda_{k}$ in $\Lambda_{k-1}$: all subunit $\Lambda_{k}$ are represented by an edge, the one represented by blue edge are the  subunit $\Lambda_{k}$ corresponding to value of $i_k=0, 1, 2, \cdots, 2m+2$.
 For example, in the  FSFT of generation $2$ shown in  Figure \ref{level_nodes}, the subunit  represented by blue dotted line,   which is a subunit of level $2$,  is labeled by a sequence $\{2, 4 \}$.  \par
  For convenience, we label the two hubs of subunit $\Lambda_k$ as $A_k, B_k$  and building mapping between hubs of $\Lambda_{k-1}$ and hubs of $\Lambda_{k}$ as shown in Figure \ref{lael_node}.  The hub of $\Lambda_{k-1}$ labeled as $A_{k-1}$ is also a hub of $\Lambda_{k}$ labeled as $A_k$  while $i_k=0, 1, 2, \cdots, m$. The hub of $\Lambda_{k-1}$ labeled as $B_{k-1}$ is also a hub of $\Lambda_{k}$ labeled as $B_k$ while $i_k=m+1, m+2, \cdots, 2m+1$.
For any $k\geq0$, let
\begin{equation}\label{Def_SK}
  \mathcal{S}^{(k)}\equiv \left( \begin{array}{c} S_{A_k}\\S_{B_k} \end{array} \right)
\end{equation}
\begin{equation}\label{Def_WK}
  \mathcal{W}^{(k)}\equiv \left( \begin{array}{c} W_{A_k}\\W_{B_k} \end{array} \right)
\end{equation}
As derived in in \ref{Pof_Rec_sk}, \ref{Pof_Rec_wk}, we obtain the following results.
\begin{lemm}   \label{Rec_Sk}  
 For any  $k>0$ , $\mathcal{S}^{(k)}$  satisfy the following recursion relations
\begin{equation}\label{SK}
  \mathcal{S}^{(k)}=\mathcal{M}_{i_k}\mathcal{S}^{(k-1)}+\mathcal{V}_{i_k}^k, \quad  i_k=0, 1, 2, \cdots, 2m+1
\end{equation}
where
\begin{equation} \label{MV0}
\mathcal{M}_{0}=
\left(                 
  \begin{array}{cc}
   1 & 0 \\
   1/2 & 1/2
  \end{array}
\right),    \quad              
\mathcal{V}_{0}^k=\left( \begin{array}{c} 0 \\ -2^{t-k}(2m+2)^{t-k} \end{array} \right)
\end{equation}
\begin{equation} \label{MV1_m}
\mathcal{M}_{i_k}=
\left(                 
  \begin{array}{cc}
   1 & 0 \\
   1 & 0
  \end{array}
\right),     \quad  \mathcal{V}_{i_k}^k=\left( \begin{array}{c} 0\\ \eta_k \end{array} \right),  \quad i_k=1, 2,\cdots, m               
\end{equation}
\begin{equation} \label{MVm1_2m}
\mathcal{M}_{i_k}=
\left(                 
  \begin{array}{cc}
   0 & 1 \\
   0 & 1
  \end{array}
\right),     \quad
 \mathcal{V}_{i_k}^k=\left( \begin{array}{c}\eta_k \\ 0\end{array} \right),  \quad i_k=m+1, m+2,\cdots, 2m
\end{equation}
\begin{equation} \label{MV2m1}
\mathcal{M}_{2m+1}=
\left(                 
  \begin{array}{cc}
   1/2 & 1/2 \\
   0 & 1
  \end{array}
\right),   \quad               
\mathcal{V}_{2m+1}^k=\left( \begin{array}{c}  -2^{t-k}(2m+2)^{t-k}\\0 \end{array} \right)                 
\end{equation}
and  $\eta_k=2^{t-k}[(2m+2)^{t}-(2m+2)^{t-k}]$.
\end{lemm}
\begin{lemm} \label{Rec_Wk}  
 For any  $k>0$ , $\mathcal{W}^{(k)}$  satisfy the following recursion relations
\begin{equation}\label{WK}
  \mathcal{W}^{(k)}=\mathcal{M}_{i_k}\mathcal{W}^{(k-1)}+\mathcal{U}_{i_k}^k, \quad  i_k=0, 1, 2, \cdots, 2m+1
\end{equation}
where $\mathcal{M}_{i_k}(i_k=0, 1, 2, \cdots, 2m+1)$ are given by Eqs.(\ref{MV0}),(\ref{MV1_m}),(\ref{MVm1_2m}), (\ref{MV2m1}) and  $\mathcal{U}_{i_k}(i_k=0, 1, 2, \cdots, 2m+1)$  are given by
\begin{equation} \label{U02m1}
\mathcal{U}_{0}^k=\left( \begin{array}{c} 0 \\ -{2^{t-k}}(2m+2)^{-k} \end{array} \right), \quad
\mathcal{U}_{2m+1}^k=\left( \begin{array}{c}  -{2^{t-k}}(2m+2)^{-k}\\0 \end{array} \right)
\end{equation}
\begin{equation} \label{U1_m}
 \mathcal{U}_{i_k}^k=\left( \begin{array}{c} 0 \\  2^{t-k}[1-(2m+2)^{-k}]\end{array} \right),  \quad i_k=1, 2,\cdots, m          
\end{equation}
\begin{equation} \label{Um1_2m}
 \mathcal{U}_{i_k}^k=\left( \begin{array}{c} 2^{t-k}[1-(2m+2)^{-k}] \\ 0\end{array} \right),  \quad i_k=m+1, m+2,\cdots, 2m
\end{equation}
\end{lemm}
Using equation (\ref{SK}) repeatedly, we obtain
\begin{eqnarray}\label{SK0}
  \mathcal{S}^{(t)}&=&\mathcal{M}_{i_t}\mathcal{S}^{(t-1)}+\mathcal{V}_{i_t}^t  \nonumber \\
   &=&\mathcal{M}_{i_t}[\mathcal{M}_{i_{t-1}}\mathcal{S}^{(t-2)}+\mathcal{V}_{i_{t-1}}^{t-1}]+\mathcal{V}_{i_t}^t \nonumber \\
   &=& \mathcal{M}_{i_t}\mathcal{M}_{i_{t-1}}\mathcal{S}^{(t-2)}+\mathcal{M}_{i_t}\mathcal{V}_{i_{t-1}}^{t-1}+\mathcal{V}_{i_t}^t \nonumber \\
   &=& \cdots \nonumber \\
   &=&\mathcal{M}_{i_t}\mathcal{M}_{i_{t-1}}\cdots \mathcal{M}_{i_1}\mathcal{S}^{(0)}
   +\sum_{l=1}^{t-1}\mathcal{M}_{i_t}\mathcal{M}_{i_{t-1}}\cdots \mathcal{M}_{i_{l+1}}\mathcal{V}_{i_{l}}^{l}+\mathcal{V}_{i_t}^t
\end{eqnarray}
Similarity, using equation (\ref{WK}) repeatedly, we obtain
\begin{eqnarray}\label{WK0}
  \mathcal{W}^{(t)}
   &=&\mathcal{M}_{i_t}\mathcal{M}_{i_{t-1}}\cdots \mathcal{M}_{i_1}\mathcal{W}^{(0)}
   +\sum_{l=1}^{t-1}\mathcal{M}_{i_t}\mathcal{M}_{i_{t-1}}\cdots \mathcal{M}_{i_{l+1}}\mathcal{U}_{i_{l}}^{l}+\mathcal{U}_{i_t}^t
\end{eqnarray}
 As for  $\mathcal{S}^{(0)}$ and $\mathcal{W}^{(0)}$, it is easy to know
 \begin{eqnarray}
 \mathcal{S}^{(0)}=(S_{A_0}, S_{B_0})^T=S_{A_0}(1, 1)^T  \label{S0} \\
  \mathcal{W}^{(0)}=(W_{A_0}, W_{B_0})^T=W_{A_0}(1, 1)^T \label{W0}
  \end{eqnarray}
  where $(x, y)^T$ is the transpose of vector  $(x, y)$  and $S_{A_0}$ and $W_{A_0}$ are $S_x$ and $T_x$ for nodes of level $0$ respectively, which have  been derived in \ref{Der_SW0}. \par
 Noticing that any edge of $G(t)$ is a subunit of level $t$, its two end nodes are just its two hubs.  If we know the label  sequence $\{i_1, i_2, ..., i_{t} \}$ for any edge of $G(t)$, we can exactly calculate $\mathcal{S}^{(t)}$  and $\mathcal{W}^{(t)}$ for its two end nodes. Hence, we can derive the expression of $S_x$  and $W_x$ for any node $x$ of $G(t)$.
 \subsection{Exact calculation of $\Sigma$ }
\label{SSWP}
We find that
\begin{equation} \label{xigma}
  \Sigma=\sum_{u\in \Omega}(\pi(u)\sum_{x\in \Omega}L_{xu})=\frac{1}{2E_t}\sum_{u\in \Omega}(d_uS_u)
 \end{equation}
$\sum_{u\in \Omega}(d_uS_u)$ is just the summation of  $S_x$ for the two end nodes of every edges of $G(t)$(Note: for node $x$ which is the intersection of $n$ edges,  $S_x$  will be counted $n$ times).   Because any edge of $G(t)$ is a subunit of level $t$, which is in one to one correspondence with a sequence  $\{i_1,\cdots,i_t\}$, its two end nodes are also its two hubs labeled as $A_t, B_t$.  Thus
\begin{equation} \label{Uxigma}
  \sum_{u\in \Omega}(d_uS_u)=\sum\left(\sum_{\{i_1,\cdots,i_t\}}{\mathcal{S}^{(t)}}\right)
\end{equation}
for the right side of the equation, the second summation is run over all the subunits of  level $t$(i.e., let $\{i_1,\cdots,i_t\}$ run over all the possible values), the first summation is just add the two entries of $\sum_{\{i_1,\cdots,i_t\}} \mathcal{S}^{(t)}$ together.\par
 Making use of the following identity
\[
\sum_{\{i_1,\cdots,i_t\}} \sum_{l=1}^{t-1} = \sum_{l=1}^{t-1} \sum_{\{i_1,\cdots,i_t\}},
\]
and define
\begin{equation} \label{MTotal}
\mathcal{M}_{tot} =\sum_{i=0}^{2m+1} \mathcal{M}_{i}
\end{equation}
\begin{equation} \label{VTotal}
\mathcal{V}_{tot}^l = \sum_{i=0}^{2m+1} \mathcal{V}_{i}^{l}
\end{equation}
 we have
\begin{eqnarray}\label{SuoV}
  \sum_{\{i_1,\cdots,i_t\}}\mathcal{M}_{i_t}\mathcal{M}_{i_{t-1}}\cdots \mathcal{M}_{i_{l+1}}\mathcal{V}_{i_{l}}^{l}= (2m+2)^{l-1}\mathcal{M}_{tot}^{t-l}\mathcal{V}_{tot}^l
\end{eqnarray}
Thus
\begin{eqnarray}\label{SST}
  \hspace{-20mm} \sum_{\{i_1,\cdots,i_t\}} \mathcal{S}^{(t)}&=& \sum_{\{i_1,\cdots,i_t\}}\left[ \mathcal{M}_{i_t}\mathcal{M}_{i_{t-1}}\cdots \mathcal{M}_{i_1}\mathcal{S}^{(0)}  
+\sum_{l=1}^{t-1}\mathcal{M}_{i_t}\mathcal{M}_{i_{t-1}}\cdots \mathcal{M}_{i_{l+1}}\mathcal{V}_{i_{l}}^{l}+\mathcal{V}_{i_t}^t \right] \nonumber \\
 \hspace{-20mm}   &=&\mathcal{M}_{tot}^{t}\mathcal{S}^{(0)}+\sum_{l=1}^{t-1}(2m+2)^{l-1}\mathcal{M}_{tot}^{t-l}\mathcal{V}_{tot}^l+(2m+2)^{t-1}\mathcal{V}_{tot}^t  \nonumber \\
  \hspace{-20mm}  &=& \mathcal{M}_{tot}^{t}\mathcal{S}^{(0)}+\sum_{l=1}^{t}(2m+2)^{l-1}\mathcal{M}_{tot}^{t-l}\mathcal{V}_{tot}^l
\end{eqnarray}
 Substituting  $\mathcal{M}_{i}$ with Eq.(\ref{MV0}), (\ref{MV1_m}), (\ref{MVm1_2m}) and (\ref{MV2m1}) in Eq. (\ref{MTotal}), and orthogonal decomposing $\mathcal{M}_{tot}$, we obtain
 \begin{eqnarray}
\mathcal{M}_{total}&=&
\left(                 
  \begin{array}{cc}
   m+3/2 & m \\
   m & m+3/2
  \end{array}
\right) \nonumber \\
&=&                  
\left(                 
  \begin{array}{cc}
\frac{\sqrt{2}}{2} & -\frac{\sqrt{2}}{2} \\
   \frac{\sqrt{2}}{2} & \frac{\sqrt{2}}{2}
  \end{array}
\right)                 
\left(                 
  \begin{array}{cc}
   2m+2 & 0 \\
   0 & 1
  \end{array}
\right)                 
\left(                 
  \begin{array}{cc}
\frac{\sqrt{2}}{2} & \frac{\sqrt{2}}{2} \\
   -\frac{\sqrt{2}}{2} & \frac{\sqrt{2}}{2}
  \end{array}
\right)    \label{MTD}             
\end{eqnarray}
Therefore
\begin{equation} \label{MTK}
\mathcal{M}_{total}^k=
\left(                 
  \begin{array}{cc}
\frac{\sqrt{2}}{2} & -\frac{\sqrt{2}}{2} \\
   \frac{\sqrt{2}}{2} & \frac{\sqrt{2}}{2}
  \end{array}
\right)                 
\left(                 
  \begin{array}{cc}
   (2m+2)^k & 0 \\
   0 & 1
  \end{array}
\right)                 
\left(                 
  \begin{array}{cc}
\frac{\sqrt{2}}{2} & \frac{\sqrt{2}}{2} \\
   -\frac{\sqrt{2}}{2} & \frac{\sqrt{2}}{2}
  \end{array}
\right)                 
\end{equation}
 Similarity, we get
 \begin{equation} \label{VT}
\mathcal{V}_{tot}^l=\left[m2^{t-k}(2m+2)^t-2^t(m+1)(2m+2)^{t-k}\right]\left( \begin{array}{c} 1 \\ 1 \end{array} \right)
\end{equation}
 Thus
 \begin{eqnarray}\label{MTS0}
  \mathcal{M}_{tot}^{t}\mathcal{S}^{(0)}&=&
  \left(                 
  \begin{array}{cc}
\frac{\sqrt{2}}{2} & -\frac{\sqrt{2}}{2} \\
   \frac{\sqrt{2}}{2} & \frac{\sqrt{2}}{2}
  \end{array}
\right)                 
\left(                 
  \begin{array}{cc}
   (2m+2)^t & 0 \\
   0 & 1
  \end{array}
\right)                 
\!\left( \!                
  \begin{array}{cc}
\frac{\sqrt{2}}{2} & \frac{\sqrt{2}}{2} \\
   -\frac{\sqrt{2}}{2} & \frac{\sqrt{2}}{2}
  \end{array}
\!\right)\!                 
\!\left(\! \begin{array}{c} 1\\1\end{array} \!\right)\!S_{A_0}
 \nonumber \\
   &=&\left(                 
  \begin{array}{cc}
\frac{\sqrt{2}}{2} & -\frac{\sqrt{2}}{2} \\
   \frac{\sqrt{2}}{2} & \frac{\sqrt{2}}{2}
  \end{array}
\right)                 
\left(                 
  \begin{array}{cc}
   (2m+2)^t & 0 \\
   0 & 1
  \end{array}
\right)                 
 \left( \begin{array}{c} \sqrt{2}\\0 \end{array} \right) S_{A_0}
\nonumber \\
   &=&\left(                 
  \begin{array}{cc}
\frac{\sqrt{2}}{2} & -\frac{\sqrt{2}}{2} \\
   \frac{\sqrt{2}}{2} & \frac{\sqrt{2}}{2}
  \end{array}
\right)                 
 \left( \begin{array}{c} (2m+2)^t \sqrt{2}\\0 \end{array} \right) S_{A_0} \nonumber \\
   &=& (2m+2)^t S_{A_0}
     \left( \begin{array}{c} 1\\ 1 \end{array} \right)
\end{eqnarray}
and
\begin{eqnarray}\label{MTVT}
  \hspace{-15mm}&&\sum_{l=1}^{t}(2m+2)^{l-1}\mathcal{M}_{tot}^{t-l}\mathcal{V}_{tot}^l\nonumber \\
 \hspace{-15mm} &=&   \sum_{l=1}^{t}(2m+2)^{l-1}
  \left\{
    \left(                 
  \begin{array}{cc}
\frac{\sqrt{2}}{2} & -\frac{\sqrt{2}}{2} \\
   \frac{\sqrt{2}}{2} & \frac{\sqrt{2}}{2}
  \end{array}
\right)                 
\left(                 
  \begin{array}{cc}
   (2m+2)^{t-l} & 0 \\
   0 & 1
  \end{array}
\right)                 
  \right.       \nonumber \\         
\hspace{-15mm}&& \cdot \left.
\left(                 
  \begin{array}{cc}
\frac{\sqrt{2}}{2} & \frac{\sqrt{2}}{2} \\
   -\frac{\sqrt{2}}{2} & \frac{\sqrt{2}}{2}
  \end{array}
\right)          
\left( \begin{array}{c} 1 \\ 1 \end{array} \right)\left[m2^{t-l}(2m+2)^t-2^t(m+1)(2m+2)^{t-l}\right]
\right\}
 \nonumber \\
\hspace{-15mm}  &=&   \sum_{l=1}^{t}(2m+2)^{t-1}\left[m2^{t-l}(2m+2)^t-2^t(m+1)(2m+2)^{t-l}\right]\left( \begin{array}{c} 1 \\ 1 \end{array} \right)
  \nonumber \\
\hspace{-15mm}  &=&\left[   m(2m+2)^{2t-1}\sum_{l=1}^{t}2^{t-l}-(m+1)(2m+2)^{t-1}\sum_{l=1}^{t}(4m+4)^{t-l}\right]\left( \begin{array}{c} 1 \\ 1 \end{array} \right)
 \nonumber \\
\hspace{-15mm}  &=&\left[m(2m+2)^{2t-1}(2^t-1)-(m+1)(2m+2)^{t-1}\frac{(4m+4)^t-1}{4m+3}\right]\left( \begin{array}{c} 1 \\ 1 \end{array} \right)
\end{eqnarray}
Inserting Eqs. (\ref{MTS0}), (\ref{MTVT}) and (\ref{SA}) into Eq.(\ref{SST}), 
we obtain
\begin{eqnarray}\label{SSTC}
  \hspace{-25mm}&&\sum_{\{i_1,\cdots,i_t\}} \mathcal{S}^{(t)}
      =\mathcal{M}_{tot}^{t}\mathcal{S}^{(0)}+\sum_{l=1}^{t}(2m+2)^{l-1}\mathcal{M}_{tot}^{t-l}\mathcal{V}_{tot}^l\nonumber \\
 \hspace{-25mm}&\!=\!&\!\left\{\!(2m\!+\!2)^{2t}\!+\!(2^t\!-\!1)(3m\!+\!1)(2m\!+\!2)^{2t-1}\!-\!\frac{m\!+\!1}{4m\!+\!3}(2m\!+\!2)^{2t-1}[(4m\!+\!4)^t\!-\!1]\!\right\}\!\left(\! \begin{array}{c} 1 \\ 1 \end{array}\! \right)\!
\end{eqnarray}
Replacing $\sum_{\{i_1,\cdots,i_t\}}\mathcal{S}^{(t)}$ with Eq.(\ref{SSTC}) in Eq. (\ref{Uxigma}),
we obtain
\begin{eqnarray}\label{SAPL}
 \hspace{-18mm}\Sigma
 &=&\frac{1}{2E_t}\sum_{u\in \Omega}\left(d_u S_u\right)\nonumber \\
 \hspace{-18mm}&=&(3m\!+\!1\!-\!\frac{m+1}{4m+3})2^t(2m+2)^{t-1}\!-\!\frac{m-1}{2m+2}(2m+2)^{t}\!+\!\frac{m+1}{(4m+3)(2m+2)}
\end{eqnarray}
\subsection{Examples}
\label{sec:example}
According to the methods presented in \ref{Met_SW}   and \ref{SSWP},  we can calculate  $T_x$ and $D_x$ for any node $x$ of $G(t)$.
We don't intend to calculate these quantities for every  node of $G(t)$ because the total number of nodes increasing rapidly with the growth of $t$. In order to explain our methods,   we calculate the  MTT or MDT for nodes of level $0$ which are labeled as $A_0$ and $B_0$,\par
Inserting Eqs.(\ref{SA} ), (\ref{WA} ) and (\ref {SAPL} )  into Eq.(\ref{MTT}) and Eq.(\ref{MDT}), we obtain the MTT  and MDT for nodes $A_0$ and $B_0$.
\begin{eqnarray} \label{TA0}
T_{B_0}=T_{A_0}
&=&S_{A_0}+N_t W_{A_0}-\Sigma  \nonumber  \\
&=&\frac{2m+2}{4m+3}2^t(2m+2)^t+\frac{2m+1}{2m+2}2^t-\frac{4m^2+4m+1}{(4m+3)(2m+2)}
\end{eqnarray}
and
\begin{eqnarray} \label{DA0}
D_{B_0}&=&D_{A_0}=S_{A_0}+\Sigma-N_t W_{A_0}  \nonumber  \\
   &=&\frac{1}{m+1}(2m+2)^t+\frac{6m^2+6m+1}{(4m+3)(m+1)}2^t(2m+2)^t \nonumber  \\
   & &-\frac{2m+1}{2m+2}2^t+\frac{4m^2+4m+1}{(4m+3)(2m+2)}
\end{eqnarray}
 We have conducted numerical simulation to test the results we have just derived, the results just derived are consistent with those obtained by numerical simulation.
\section{Effect of node position on trapping efficiency  for random walks on FSFT}
\label{sec:Eff_MTT}
In these section, using the MTT  as the measure of trapping efficiency, we compare the trapping efficiency (i.e., the MTT) among all the nodes of FSFT and find the best trapping sites(i.e., nodes which have the minimum MTT) and the  worst trapping sites(i.e., nodes which have the maximum MTT). \par
In order to compare the MTT for nodes of different levels, we  derive the relations of $T_x$ for nodes of level $k$ and that for nodes of level $k+1$, and then compare $T_x$ for nodes of adjacent level.\par
\begin{figure}
\begin{center}
\includegraphics[scale=0.4]{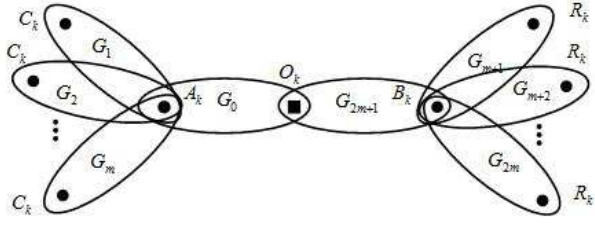}
\caption{Construction for subunit $\Lambda_{k}$: it is  composed of $2m+2$  subunits of level $k$ labeled as  $G_0$, $G_1$, $\cdots$, $G_{2m+1}$. Nodes labeled as $A_k$, $B_k$, $O_k$ , $C_k$ and $R_k$ are all hubs of the $2m+2$  subunits.}
\label{Relation}
\end{center}
\end{figure}

Considering any subunit of level $k$ as shown in Figure \ref{Relation},  it is  composed of $2m+2$  subunits of level $k+1$ (black oval with solid line).  its two hubs (i.e., $A_k$ and $B_k$) are the only two nodes of level $k$,  its nodes of level $k+1$ are hubs of its $2m+2$  subunits of level $k+1$(i.e., $A_k$, $B_k$, $O_k$ , $C_k$ and $R_k$). Assuming $T_x$ for node  of level $k$(i.e., $T_{A_k}$, $T_{B_k}$) are known, we will analyze $T_x$ for node $x$ of level $k+1$(i.e., $O_k$,  $C_k$ and $R_k$).\par
For any $k\geq 0$,  it is easy to obtain the following equation due to  Eqs. (\ref{MTT}), (\ref{SCi}),  (\ref{SCm1_2m1}), (\ref{SO}),  (\ref{WCi}),  (\ref{WCm1_2m1}) and (\ref{WO}).
\begin{eqnarray} \label{TOk}
T_{O_{k}}&=&S_{O_{k}}-\Sigma+N_t W_{O_{k}}  \nonumber  \\
&=&\frac{1}{2}(S_{A_{k}}+S_{B_{k}})-2^{t-k-1}(2m+2)^{t-k-1}-\Sigma \nonumber  \\
&&+N_t\cdot \left[\frac{1}{2}(W_{A_{k}}+W_{B_{k}})-2^{t-k-1}(2m+2)^{-k-1}\right] \nonumber  \\
&=&\frac{1}{2}(T_{A_{k}}+T_{B_{k}})-2^{t-k}(2m+2)^{t-k-1}-2^{t-k-1}(2m+2)^{-k-1}
\end{eqnarray}
and
\begin{eqnarray} \label{TCk}
\hspace{-15mm}T_{C_{k}}&=&S_{C_{k}}-\Sigma +N_t W_{C_{k}} \nonumber  \\
\hspace{-15mm}&=&S_{A_{k}}+2^{t-k-1}[(2m+2)^t-(2m+2)^{t-k-1}]-\Sigma \nonumber  \\
\hspace{-15mm}&&+N_t\cdot \left\{W_{A_{k}}-2^{t-k-1}[1-(2m+2)^{-k-1}]\right\} \nonumber  \\
\hspace{-15mm}&=&T_{A_{k}}+2^{t-k}[(2m+2)^t-(2m+2)^{t-k-1}]+2^{t-k-1}[1-(2m+2)^{-k-1}]
\end{eqnarray}
\begin{eqnarray} \label{TRk}
\hspace{-15mm}T_{R_{k}}
&=&T_{B_{k}}+2^{t-k}[(2m+2)^t-(2m+2)^{t-k-1}]+2^{t-k-1}[1-(2m+2)^{-k-1}]
\end{eqnarray}
Note that $T_{A_0}=T_{B_0}$ and let $k=0$ in Eqs.(\ref{TOk}), (\ref{TCk}) and (\ref{TRk}), we find
\begin{eqnarray} \label{TCom0}
   T_{O_{0}}< T_{A_0}=T_{B_0}< T_{C_{0}}=T_{R_{0}}
\end{eqnarray}
For $k\geq1$, it easy to derive from Eqs.(\ref{TCk}) and (\ref{TRk}) that
 \begin{equation}
     T_{C_{k}}>T_{A_k} \qquad and \qquad T_{R_{k}}> T_{B_k}
       \label{ComTAC}
 \end{equation}
 As proved in \ref{Pro_ComTOk}, we  find Eq.(\ref{ComTOk}) holds for $k\geq1$.
 \begin{equation}
      min\{T_{A_k},   T_{B_k}\}< T_{O_{k}}< max\{T_{A_k},   T_{B_k}\}
       \label{ComTOk}
\end{equation}
Therefore, for $k\geq1$
   \begin{equation}
      min\{T_{A_k},   T_{B_k}\}=  min\{T_{A_k},   T_{B_k}, T_{O_{k}}, T_{C_{k}}, T_{R_{k}}\}
       \label{ComTk}
\end{equation}
Let $ \Omega_k$ denote set for nodes of level $k$ and note that $A_k$ and $B_k$ are the only two nodes of level $k$ in $\Lambda_k$,   $\{{A_k},   {B_k}, {O_{k}}, {C_{k}}, {R_{k}}\}$ represents all nodes of level $k+1$  in $\Lambda_k$,
   Eq.(\ref{ComTk}) implies
 \begin{equation}\label{Tmink}
       min \{T_x: x\in\Omega_k\}=  min \{T_x: x\in\Omega_{k+1}\}, \qquad  k\geq1
 \end{equation}
But Eqs.(\ref{TCom0})  shows
     \begin{equation}\label{Tmin1}
       T_{O_{0}}=min\{T_x: x \in \Omega_1\}<min\{T_x: x \in \Omega_0\}
     \end{equation}
Thus
\begin{equation}\label{Tmin}
       T_{O_{0}}=min\{T_x: x \in \Omega_1\}=min\{T_x: x \in \Omega_t\}=min\{T_x: x \in \Omega\}
 \end{equation}
  Let $k=0$ and replacing $T_{A_0}$ and $T_{B_0}$ with Eq.(\ref{TA0}) in Eq.(\ref{TOk}),  the minimum of the MTT is
\begin{eqnarray} \label{TO0}
T_{O_{0}}
&=&\frac{4m^2\!+\!4m\!+\!1}{(4m\!+\!3)(2m\!+\!2)}2^t(2m\!+\!2)^t\!+\!\frac{4m\!+\!1}{4m\!+\!4}2^t\!-\!\frac{4m^2\!+\!4m\!+\!1}{(4m\!+\!3)(2m\!+\!2)}
\end{eqnarray}
The result of $T_{O_{0}}$   is consistent with that  derived in Ref. \cite{ZhLi11} which shows the  correctness of our methods.\par
As for the maximum of MTT, we can derive from Eqs. (\ref{TCom0}), (\ref{ComTAC}) and (\ref{ComTOk}) that
  \begin{equation}\label{Tmaxk}
       max \{T_x: x\in\Omega_k\}< max \{T_x: x\in\Omega_{k+1}\}, \qquad  k\geq0
 \end{equation}
We can also derive from Eqs. (\ref{TCk})  and (\ref{TRk}) that  the nodes with max MTT among nodes of level $k+1$  are the nodes which  directly connected to the nodes of level $k$ with max MTT among nodes of level $k$. According to the structure of FSFT, the nodes with max MTT among all nodes of level $k+1$ are  the nodes which are farthest from nodes of level $0$. Let $T_{max}^k$ denote the maximum of MTT among nodes of level $k$, it is easy to know that $$T_{max}^0=T_{A_0}$$
 and for $k\geq$, We can also obtain from Eqs.(\ref{TCk}) and (\ref{TRk})
\begin{eqnarray} \label{Tmaxk1}
\hspace{-15mm}T_{max}^{k+1}&=&T_{max}^{k}+2^{t-k}[(2m+2)^t-(2m+2)^{t-k-1}]+2^{t-k-1}[1-(2m+2)^{-k-1}]
\end{eqnarray}
Using Eq.(\ref{Tmaxk1}) repeatedly and replacing $T_{max}^{0}$ with Eq.(\ref{TA0}), we obtain
\begin{eqnarray} \label{Tmax}
\hspace{-15mm}T_{max}^{t}&=&T_{max}^{0}\!+\!\sum_{k=1}^t\left\{2^{t-k+1}[(2m+2)^t\!-\!(2m+2)^{t-k}]\!+\!2^{t-k}[1\!-\!(2m+2)^{-k}]\right\} \nonumber  \\
\hspace{-15mm}&=&T_{max}^{0}\!+\!(2m\!+\!2)^t\sum_{k=1}^t2^{t\!-\!k\!+\!1}\!-\!2\sum_{k=1}^t(4m\!+\!4)^{t\!-\!k}\!+\!\sum_{k=1}^t2^{t\!-\!k}\!-\!2^t\sum_{k=1}^t(4m\!+\!4)^{\!-\!k} \nonumber  \\
\hspace{-15mm}&=&(2m+2)^t\!\cdot\!2^t\cdot\frac{10m+6}{4m+3}\!-\!2(2m+2)^t\!+\!2^{t}\frac{16m^2+22m+7}{(4m+3)(2m+2)}\nonumber \\
\hspace{-15mm}& &\!+\!\frac{1}{(2m+2)^{t}(4m+3)}\!-\!\frac{6m+3}{4m+3}\!+\!\frac{2}{4m+3}
\end{eqnarray}
Because all nodes of $G(t)$ belong to level $t$, $T_{max}^{t}$ is the maximum of MTT among all nodes of FSFT $G(t)$.
Comparing $T_{max}^{t}$  with $T_{O_0}$ shown in Eq. (\ref{TO0}), and let $t\rightarrow\infty$, we obtain
 \begin{equation} \label{Com_MTT}
\frac{T_{max}^{t}}{T_{O_0}}\approx \frac{20m^2+32m+12}{4m^2+4m+1}>5
\end{equation}
which shows that  maximum  of MTT is almost $\frac{20m^2+32m+12}{4m^2+4m+1}$ times of the minimum of MTT, thus the position of target node has big effect on trapping efficiency.

\section{Effect of node position on diffusion efficiency for random walks on FSFT}
\label{sec:Eff_diff}
In these section, using the MDT  as the measure of trapping efficiency, we compare the trapping efficiency (i.e., the MDT) among all the nodes of FSFT and find the best trapping sites(i.e., nodes which have the minimum MTT) and the  worst trapping sites(i.e., nodes which have the maximum MDT). \par
Similarity to the analysis of trapping efficiency, we first derive the relations of $D_x$ for nodes of level $k$ and that for nodes of level $k+1$, and then compare $D_x$ for nodes of adjacent level, finally, we find and compare the minimum and maximum of MDT among all nodes of FSFT.\par
Considering any subunit of level $k (k\geq 0)$ as shown in Figure \ref{Relation},  it is easy to obtain the following equation due to  Eqs. (\ref{MDT}), (\ref{SCi}),  (\ref{SCm1_2m1}), (\ref{SO}),  (\ref{WCi}),  (\ref{WCm1_2m1}) and (\ref{WO}).
\begin{eqnarray} \label{DOk}
D_{O_{k}}
&=&\frac{1}{2}(D_{A_{k}}+D_{B_{k}})+2^{t-k-1}(2m+2)^{-k-1}
\end{eqnarray}
 \begin{eqnarray} \label{DCk}
D_{C_{k}}
&=&D_{A_{k}}-2^{t-k-1}[1-(2m+2)^{-k-1}]
\end{eqnarray}
and
\begin{eqnarray} \label{DRk}
D_{R_{k}}
&=&D_{B_{k}}-2^{t-k-1}[1-(2m+2)^{-k-1}]
\end{eqnarray}
Note that $D_{A_0}=D_{B_0}$ and let $k=0$ in Eqs.(\ref{DOk}), (\ref{DCk}) and (\ref{DRk}), we find
\begin{eqnarray} \label{DCom0}
   D_{O_{0}}> D_{A_0}=D_{B_0}> D_{C_{0}}=D_{R_{0}}
\end{eqnarray}
For $k\geq 1$, it easy to derive from Eqs.(\ref{DCk}) and (\ref{DRk}) that
 \begin{equation}
     D_{C_{k}}<D_{A_k} \qquad and \qquad D_{R_{k}}< D_{B_k}
       \label{ComDAC}
 \end{equation}
 As proved in \ref{Pro_ComDOk}, we  find Eq.(\ref{ComDOk}) holds for $k\geq1$.
  \begin{equation}
     min\{D_{A_k},   D_{B_k}\}< D_{O_{k}}< max\{D_{A_k},   D_{B_k}\}
       \label{ComDOk}
   \end{equation}
Therefore, for $k\geq1$, we have
  \begin{equation}
     max\{D_{A_k},   D_{B_k}\}=max\{D_{A_k},  D_{B_k}, D_{O_{k}}, D_{C_{k}}, D_{R_{k}}\}
       \label{maxDk}
   \end{equation}
Because $A_k$ and $B_k$ are the only two nodes of level $k$ in $\Lambda_k$ and  $\{{A_k},  {B_k}, {O_{k}}, {C_{k}}, {R_{k}}\}$ represents all nodes of level $k+1$  in $\Lambda_k$,
Eqs.(\ref{maxDk}) implies that
 \begin{equation}\label{Dmaxk}
       max \{D_x: x\in\Omega_k\}=  max \{D_x: x\in\Omega_{k+1}\}, \qquad  k\geq1
 \end{equation}
But Eqs.(\ref{DCom0})  lead to
     \begin{equation}\label{Dmin1}
       D_{O_{0}}=max\{D_x: x \in \Omega_1\}>min\{D_x: x \in \Omega_0\}
     \end{equation}
Thus
\begin{equation}\label{Dmin}
       D_{O_{0}}=max\{D_x: x \in \Omega_1\}=max\{D_x: x \in \Omega_t\}=max\{D_x: x \in \Omega\}
 \end{equation}
  Let $k=0$ and replacing $D_{A_0}$ and $D_{B_0}$ with Eq.(\ref{DA0}) in Eq.(\ref{DOk}),  the maximum of the MDT is
\begin{eqnarray} \label{DO0}
D_{O_{0}}&=&\frac{1}{m+1}(2m+2)^t+\frac{6m^2+6m+1}{(4m+3)(m+1)}2^t(2m+2)^t \nonumber  \\
   & &-\frac{4m+1}{4m+4}2^t+\frac{4m^2+4m+1}{(4m+3)(2m+2)}
\end{eqnarray}
As for the minimum of MDT, we can derive from Eqs. (\ref{DCom0}), (\ref{ComDAC}) and (\ref{ComDOk}) that
  \begin{equation}\label{Dmink}
       min \{D_x: x\in\Omega_k\}> min \{D_x: x\in\Omega_{k+1}\}, \qquad  k\geq0
 \end{equation}
Similarity to the analysis of maximum of MTT,  we find that  the nodes with min MDT among nodes of level $k$  are just the nodes which  have max MTT among nodes of level $k$.  Let $D_{min}^k$ denote the minimum of MDT among nodes of level $k$, it is easy to know that $$T_{min}^0=D_{A_0}$$
 We can also obtain from Eqs.(\ref{DCk}) and (\ref{DRk})
\begin{eqnarray} \label{Dmink1}
\hspace{-15mm}D_{min}^{k+1}&=&D_{min}^k-2^{t-k-1}[1-(2m+2)^{-k-1}]
\end{eqnarray}
Using Eq.(\ref{Dmink1}) repeatedly and replacing $D_{min}^{0}$ with Eq.(\ref{DA0}), the minimum of MDT among all nodes of FSFT is
\begin{eqnarray} \label{Dmax}
\hspace{-15mm}D_{min}^{t}&=&D_{min}^{0}\!+\!\sum_{k=1}^t\left\{2^{t-k}[1-(2m+2)^{-k}]\right\} \nonumber  \\
\hspace{-15mm}&=&D_{min}^{0}\!-\!\sum_{k=1}^t2^{t\!-\!k\!}\!+\!2^t\sum_{k=1}^t(4m\!+\!4)^{t\!-\!k} \nonumber  \\
\hspace{-15mm}&=&\frac{(2m\!+\!2)^t}{m+1}\!+\! (2m\!+\!2)^t2^t\cdot\frac{6m^2+6m+1}{(4m+3)(m+1)} \!-\!2^{t}\frac{16m^2+22m+7}{(4m+3)(2m+2)}\nonumber \\
\hspace{-15mm}& &\!-\!\frac{1}{(2m+2)^{t}(4m+3)}\!+\!\frac{6m+3}{4m+3}
\end{eqnarray}
Comparing $D_{O_0}$ with $D_{min}^{t}$, and let $t\rightarrow\infty$, we obtain
 \begin{equation} \label{Com_MTT}
\frac{D_{O_0}}{D_{min}^{t}}\approx 1
\end{equation}
which implies that the difference between maximum  and minimum of MDT is quite small, thus the position of starting node almost has no effect on diffusion efficiency.

\section{Conclusion}
\label{sec:4}
In this paper,we study unbiased discrete random walks on FSFT. First, we  provided general methods for calculating  the mean trapping time(MTT) for any target node and the mean diffusing time(MDT)  for any source node, and then we gave some examples to explain our methods.
Finally, using the MTT and the MDT as the measures of trapping efficiency and diffusion efficiency respectively, we compare the trapping efficiency and diffusion efficiency among all nodes  of FSFT and find the best ( or worst)  trapping sites  and the best ( or worst) diffusing sites. Our results show that: the central node $O$  of FSFT is the best trapping site, but it is also the worst diffusing site, the   nodes which are the farthest nodes from the two hubs are the worst trapping sites, but they are also the best diffusion sites.  Comparing the maximum  and minimum of MTT and MDT, we found that the  maximum  and minimum of MTT have big difference, but the difference between maximum  and minimum of MDT is quite small, thus the trap's position has big effect on the trapping efficiency, but the position of starting node almost has no effect on diffusion efficiency. The methods  we present can also be used  on other self-similar trees.

\ack{
The authors are grateful to the anonymous referees for their valuable comments and suggestions. This work was supported  by
the scientific research program of Guangzhou municipal colleges and universities under Grant No. 2012A022.
}

\appendix 
\section{Proof of Lemma \ref{Rec_Sk}  }
\label{Pof_Rec_sk}
\begin{figure}
\begin{center}
\includegraphics[scale=0.5]{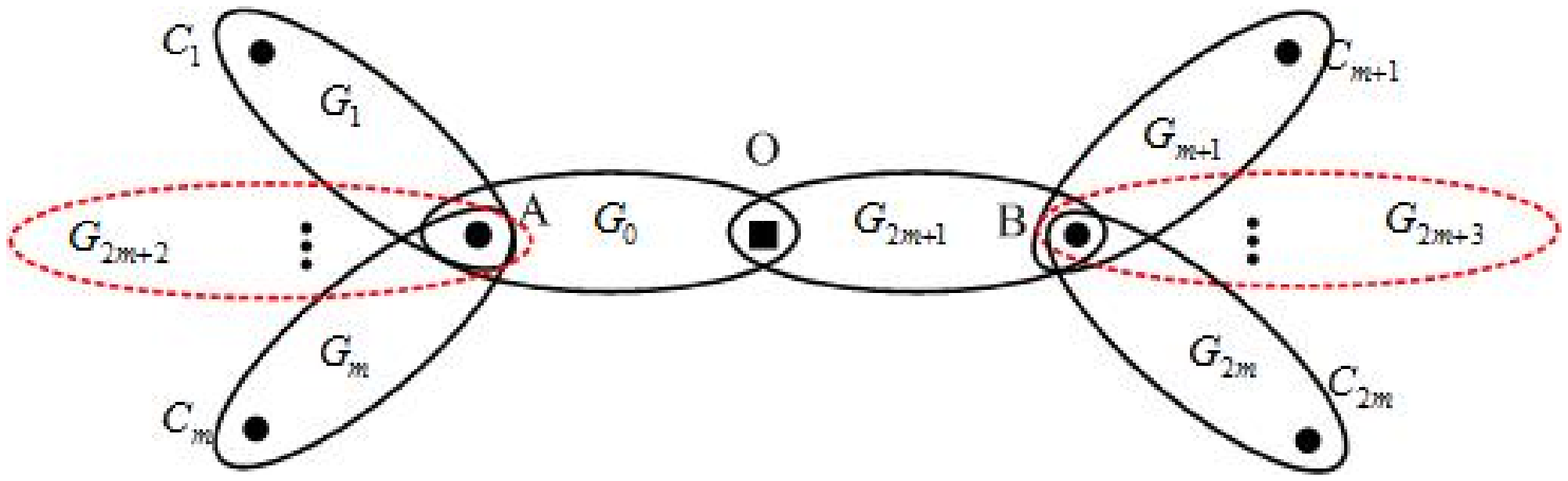}
\caption{Construction for subunit $\Lambda_{k-1}$: it is  composed of $2m+2$  subunits of level $k$ labeled as  $G_0$, $G_1$, $\cdots$, $G_{2m+1}$ (oval with solid line in the figure),  The subunit labeled as $G_{2m+2}$ and $G_{2m+3}$ denote the the rest part of the FSFT $G(t)$ except for the subunit $\Lambda_{k-1}$ (red oval with dotted line in the figure), where $G_{2m+2}$ connect with $\Lambda_{k-1}$ by nodes $A$, while $G_{2m+3}$ connect with $\Lambda_{k-1}$ by nodes $B$.}
\label{Subk_1_Rec}
\end{center}
\end{figure}

Considering any subunit of level $k-1$ as shown in Figure \ref{Subk_1_Rec}, it is  composed of $2m+2$  subunits of level $k$. It is also connect with other part of the  FSFT by the two hubs (i.e. A and B in Figure \ref{Subk_1_Rec}). In  this subunit, the two hubs are the only two nodes of level $k-1$,  its nodes of level $k$ are hubs of its $2m+2$  subunits of level $k$(i.e.$A$, $B$, $O$ and $C_i(i=1, 2, \cdots, 2m)$ in Figure \ref{Subk_1_Rec}). Assuming $S_x$ for node  of level $k-1$(i.e., $S_A$, $S_B$) is known, we will analyze $S_x$ for node $x$ of level $k$(i.e., $O$ and $C_i(i=1, 2, \cdots, 2m)$).\par
Let
\begin{equation}
S_x(i)=\sum_{y \in G_i}{L_{xy}} \qquad \qquad i=0, 1, \cdots, 2m+3
\end{equation}
where \lq\lq$y\in G_i$\rq\rq means that $y$ belongs to the nodes set of $G_i$. Thus
\begin{equation}
S_x=\sum_{i=0}^{2m+3}S_x(i)-mL_{xA}-mL_{xB}-L_{xO} \label{SX2}
\end{equation}
First, we calculate $S_{C_1}$.  $C_1$ and $A$ are the two hubs of  $G_1$ which is a subunit of level $k$, the distance between $C_1$ and $A$ is $L_{AC_1}=2^{t-k}$ . The total numbers of nodes of  $G_1$ is $N_{t-k}$. It is easy to obtain that $S_{C_1}(1)=S_A(1)$ and for node $y\in G_i(i\neq 1)$, $L_{yC_1}=L_{yA}+L_{AC_1}$.
Thus
\begin{eqnarray}
S_{C_1}&=&\sum_{i=0}^{2m+3}S_{C_1}(i)-mL_{A{C_1}}-mL_{B{C_1}}-L_{O{C_1}} \nonumber \\
&=&S_A(1)+\sum_{i\neq 1}\sum_{y \in G_i}({L_{yA}+L_{AC_1}})-(4m+2)L_{A{C_1}} \nonumber \\
&=&\sum_{i=0}^{2m+3}S_A(i)+\sum_{i\neq 1}\sum_{y \in G_i}L_{AC_1}-(4m+2)L_{A{C_1}} \nonumber \\
&=&S_A\!+\!mL_{BA}\!+\!L_{OA}\!+\![(2m+1)N_{t-k}\!+\!N_t\!-\!N_{t-k+1}\!-\!(4m+2)]L_{AC_1} \nonumber \\
&=&S_A+[N_t-N_{t-k}]L_{AC_1}\nonumber \\
&=&S_A+2^{t-k}[(2m+2)^t-(2m+2)^{t-k}]
\label{SC1}
\end{eqnarray}
Similarity, for $i=0, 1, \cdots, m$ we have
\begin{equation}\label{SCi}
  S_{C_i}=S_A+2^{t-k}[(2m+2)^t-(2m+2)^{t-k}]
\end{equation}
and for $ i=m+1, m+2, \cdots, 2m+1$, we have
\begin{equation}\label{SCm1_2m1}
  S_{C_i}=S_B+2^{t-k}[(2m+2)^t-(2m+2)^{t-k}]
\end{equation}
Now,  we calculate $S_{O}$.  Node $O$ is one hub of  $G_0$ and $G_{2m+1}$. It is easy to know $L_{AO}=L_{BO}=2^{t-k}$. We also found that
$S_{O}(0)=S_A(0)$,  and
\begin{eqnarray}
S_{O}(0)
&=&\sum_{y \in G_0}{L_{yO}} \nonumber \\
&=&\sum_{y \in G_0}({L_{yB}-L_{BO}}) \nonumber \\
&=&S_B(0)-N_{t-k}L_{BO}
\end{eqnarray}
Thus
\begin{equation}\label{SO0}
S_{O}(0)=\frac{1}{2}\left[S_A(0)+S_B(0)\right]-\frac{N_{t-k}L_{BO}}{2}
\end{equation}
 By symmetry
 \begin{equation}\label{SO2m1}
 S_{O}(2m+1)=\frac{1}{2}\left[S_A(2m+1)+S_B(2m+1)\right]-\frac{N_{t-k}L_{AO}}{2}
 \end{equation}
For any node $y\in G_i(i=1, 2, \cdots, m,2m+2)$, we have $L_{yO}=L_{yA}+L_{AO}$ and $L_{yO}=L_{yB}-L_{AO}$, which lead to  $L_{yO}=\frac{1}{2}(L_{yA}+L_{yB})$.
By symmetry,  $L_{yO}=\frac{1}{2}(L_{yA}+L_{yB})$  also holds for any node $y\in G_i(i=m+1, m+2, \cdots, 2m, 2m+3)$. Therefore, Eq. (\ref{SOi}) holds for any $i=1, 2, \cdots, 2m, 2m+2, 2m+3$.
\begin{eqnarray}
S_{O}(i)
&=&\sum_{y \in G_i}{L_{yO}} \nonumber \\
&=&\sum_{y \in G_i}\frac{1}{2}(L_{yA}+L_{yB}) \nonumber \\
&=&\frac{1}{2}\left[S_A(i)+S_B(i)\right]
\label{SOi}
\end{eqnarray}
Hence
\begin{eqnarray}
S_{O}&=&\sum_{i=0}^{2m+3}S_{O}(i)-mL_{AO}-mL_{BO} \nonumber \\
&=&\frac{1}{2}\sum_{i=0}^{2m+3}\left[S_A(i)+S_B(i)\right]-{N_{t-k}L_{BO}}-2mL_{BO} \nonumber \\
&=&\frac{1}{2}(S_A+S_B)+mL_{BA}+L_{BO}-{N_{t-k}L_{BO}}-2mL_{BO} \nonumber \\
&=&\frac{1}{2}(S_A+S_B)-2^{t-k}(2m+2)^{t-k}
\label{SO}
\end{eqnarray}
If we label the two hubs of $\Lambda_k$ as $A_k, B_k$, we have $A_{k-1}\equiv A, B_{k-1}\equiv B$. According  the following  mapping between hubs of $\Lambda_{k-1}$ and hubs of $\Lambda_{k}$, which can be derived from Figure \ref{lael_node}.
\begin{eqnarray} \label{mappingk_1}
\left \{                 
  \begin{array}{ll}
   A_{k}\equiv A_{k-1}, B_{k}\equiv O_{k-1}  & i_{k}=0 \\
   A_{k}\equiv O_{k-1}, B_{k}\equiv B_{k-1} & i_{k}=2m+1 \\
   A_{k}\equiv A_{k-1}, B_{k}\equiv C_{k-1} & i_{k}=1,2,\cdots,m \\
    A_{k}\equiv R_{k-1}, B_{k}\equiv B_{k-1} & i_{k}=m+1,m+2,\cdots,2m
  \end{array}
\right.
\end{eqnarray}
Thus, for $i_k=0$, we have
\begin{eqnarray}\label{SK0}
  \mathcal{S}^{(k)}
  \equiv \left( \begin{array}{c} S_{A}\\S_{O}\end{array} \right)
&=&\left(                 
  \begin{array}{cc}
   1 & 0 \\
   1/2 & 1/2
  \end{array}
\right)
\!\left(\! \begin{array}{c} S_{A}\\S_{O}\end{array} \!\right)\!
\!+\!\left(\! \begin{array}{c} 0\\-2^{t-k}(2m+2)^{t-k} \end{array}\! \right)\! \nonumber \\
&\equiv&\left(                 
  \begin{array}{cc}
   1 & 0 \\
   1/2 & 1/2
  \end{array}
\right)
\mathcal{S}^{(k-1)}
+\left( \begin{array}{c}  0\\-2^{t-k}(2m+2)^{t-k} \end{array} \right)
\end{eqnarray}
Therefore,  Eq. (\ref{SK}) holds  for $i_k=0$.
Similarly, we can verify that Eq. (\ref{SK}) holds  for $i_k=1, 2, \cdots, 2m+1$.
\section{Proof of Lemma \ref{Rec_Wk} }
\label{Pof_Rec_wk}
Considering any subunit of level $k-1$ as shown in Figure \ref{Subk_1_Rec}, assuming $S_x$ for node  of level $k-1$(i.e., $W_A$, $W_B$) is known, we will analyze $W_x$ for node $x$ of level $k$(i.e., $O$ and $C_i(i=1, 2, \cdots, 2m)$).\par
Let
\begin{equation}
W_x(i)=\frac{1}{2E_t}\cdot \sum_{y \in  G_i}{(L_{xy}\cdot d_y)}   \qquad \qquad i=0, 1, \cdots, 2m+3
\end{equation}
where  the degree for nodes which is the intersection of two subgraphs were counted respectively in every subgraph and the degree for the node in $\Lambda_{k-1}$ is just the summation of the degrees for the node in all the subgraph $\Lambda_{k}$. For example the degree of node $O$ in $\Lambda_{k-1}$ is just  the summation of the degree for $O$ in subgraph $G_0$ and $G_{2m+1}$. Thus
\begin{equation}
W_x=\sum_{i=0}^{2m+3}W_x(i) \label{WX2}
\end{equation}
First, we calculate $W_{C_1}$.   It is easy to obtain that $W_{C_1}(1)=W_A(1)$ and for node $y\in G_i(i\neq 1)$, $L_{yC_1}=L_{yA}+L_{AC_1}$.
Thus
\begin{eqnarray}
W_{C_1}&=&\sum_{i=0}^{2m+3}W_{C_1}(i) \nonumber \\
&=&W_A(1)+\frac{1}{2E_t}\sum_{i\neq 1}\sum_{y \in G_i}({L_{yA}+L_{AC_1}})d_y \nonumber \\
&=&\sum_{i=0}^{2m+3}W_A(i)+\frac{1}{2E_t}\sum_{i\neq 1}\sum_{y \in G_i}d_yL_{AC_1} \nonumber \\
&=&W_A +\frac{1}{2E_t}[2E_t-2E_{t-k}]L_{AC_1} \nonumber \\
&=&W_A+2^{t-k}[1-(2m+2)^{-k}]
\label{WC1}
\end{eqnarray}
Similarity, for $i=0, 1, \cdots, m$, we have
\begin{equation}\label{WCi}
   W_{C_i}=W_A+2^{t-k}[1-(2m+2)^{-k}]
\end{equation}
and for $ i=m+1, m+2, \cdots, 2m+1$, we have
\begin{equation}\label{WCm1_2m1}
  W_{C_i}=W_B+2^{t-k}[1-(2m+2)^{-k}]
\end{equation}
Now,  we calculate $W_{O}$.  Node $O$ is one hub of  $G_0$ and $G_{2m+1}$. It is easy to know $L_{AO}=L_{BO}=2^{t-k}$. We also found that
$W_{O}(0)=W_A(0)$,  and
\begin{eqnarray}
W_{O}(0)
&=&\frac{1}{2E_t}\sum_{y \in G_0}{L_{yO}}d_y \nonumber \\
&=&\frac{1}{2E_t}\sum_{y \in G_0}({L_{yB}-L_{BO}})d_y \nonumber \\
&=&W_B(0)-2^{t-k}(2m+2)^{-k}
\end{eqnarray}
Thus
\begin{equation}\label{WO0}
W_{O}(0)=\frac{1}{2}\left[W_A(0)+W_B(0)\right]-2^{t-k-1}(2m+2)^{-k}
\end{equation}
 By symmetry
 \begin{equation}\label{WO2m1}
 W_{O}(2m+1)=\frac{1}{2}\left[W_A(2m+1)+W_B(2m+1)\right]-2^{t-k-1}(2m+2)^{-k}
 \end{equation}
Note $L_{yO}=\frac{1}{2}(L_{yA}+L_{yB})$   holds for any node $y\in G_i(i=1, 2, \cdots, 2m, 2m+2, 2m+3)$. Therefore, Eq. (\ref{WOi}) holds for any $i=1, 2, \cdots, 2m, 2m+2, 2m+3$.
\begin{eqnarray}
W_{O}(i)=\frac{1}{2}\left[W_A(i)+W_B(i)\right]
\label{WOi}
\end{eqnarray}
Replacing $W_O(i)$ with Eqs.(\ref{WO0}), (\ref{WO2m1}), (\ref{WOi}) in Eq. (\ref{WX2}), we obtain
\begin{eqnarray}
W_{O}=\frac{1}{2}(W_A+W_B)-{2^{t-k}}(2m+2)^{-k}
\label{WO}
\end{eqnarray}
Similar to \ref{Pof_Rec_wk}, if we label the two hubs of $\Lambda_k$ as $A_k, B_k$  and let
\begin{equation}
  \mathcal{W}^{(k)}\equiv \left( \begin{array}{c} W_{A_k}\\W_{B_k} \end{array} \right)
\end{equation}
We can verify Eq. (\ref{WK}) holds for $i_k=0, 1, 2, \cdots, 2m+1$.
\section{Derivation of  $S_{A_0}$ and $W_{A_0}$}
\label{Der_SW0}
 $A_0$ is one of the two   nodes of level $0$ (i.e., A, B in Figure \ref{structure}), it is also one of the two hubs of $G(t)$. In order to tell the difference of $S_{A_0}$ (and $W_{A_0}$ ) for FSFT of different generation $t$,  let $S_{A_0}^t$ and $W_{A_0}^t$ denote the $S_{A_0}$ and $W_{A_0}$  in FSFT of generation $t$. It is easy to know $S_{A_0}^0=1$ and  $W_{A_0}^0=\frac{1}{2}$. For $t>1$, according  to the self-similar structure shown in Figure \ref{structure}, $S_{A_0}^t$  satisfies the following recursion relation.
\begin{equation}
S_{A_0}^t=(m+1)\cdot S_{A_0}^{t-1}+S_{A_0}^{t-1}+2^{t-1}(N_{t-1}-1) +m\cdot[S_{A_0}^{t-1}+2^{t}(N_{t-1}-1)]  \nonumber
\end{equation}
For the right side of the equation, the first item represents the summation for shortest path length between node $A$ and nodes in the subunit $G_i(t)(i=0, 1, \cdots, m)$, the second item represents the summation for shortest path length between node $A_0$ and nodes in the subunit $G_{2m+1}(t)$, the third item represents the summation for shortest path length between node $A_0$ and nodes in the subunit $G_i(t)(i=m+1, m+2, \cdots, 2m)$. Note that $N_{t-1}=(2m+1)^{t-1}+1$, thus, in FSFT of generation $t$,
\begin{eqnarray}
 S_{A_0}=S_{A_0}^t&=& (2m+2) S_{A_0}^{t-1}+(2m+1)2^{t-1}(2m+2)^{t-1}  \nonumber \\
      &=&(2m+2)\left[ (2m+2) S_{A_0}^{t-2}+(2m+1)2^{t-2}(2m+2)^{t-2} \right]\nonumber \\
      & &+(2m+1)2^{t-1}(2m+2)^{t-1}   \nonumber \\
      &=&(2m+2)^2 S_{A_0}^{t-2}+(2m+1)(2m+2)^{t-1}\left[2^{t-2}+2^{t-1} \right]\nonumber \\
      &=&\cdots \nonumber \\
      &=&(2m\!+\!2)^t S_{A_0}^{0}\!+\!(2m\!+\!1)(2m\!+\!2)^{t-1}\left[2^0\!+\!2^1\!+\!\cdots\!+\!2^{t-1} \right]  \nonumber \\
      &=&(2m+2)^t+(2m+1)(2m+2)^{t-1}(2^t-1)\label{SA}
\end{eqnarray}
Similarity, we find that  $W_{A_0}^t$  satisfies the following recursion relation.
\begin{equation}
W_{A_0}^t=\frac{1}{2m+2}\cdot\left\{(m+1)W_{A_0}^{t-1}+W_{A_0}^{t-1}+2^{t-1} +m\cdot[W_{A_0}^{t-1}+2^{t}] \right\} \nonumber
\end{equation}
Hence
\begin{eqnarray}
 W_{A_0}=W_{A_0}^t&=& W_{A_0}^{t-1}+\frac{2m+1}{2m+2}2^{t-1}  \nonumber \\
      &=&W_{A_0}^{t-2}+\frac{2m+1}{2m+2}2^{t-2}+\frac{2m+1}{2m+2}2^{t-1}  \nonumber \\
      &=&\cdots \nonumber \\
      &=&W_{A_0}^{0}+\frac{2m+1}{2m+2}\left[2^0+2^1+\cdots+2^{t-1}\right]  \nonumber \\
      &=&\frac{1}{2}+\frac{2m+1}{2m+2}(2^t-1)
      \label{WA}
\end{eqnarray}
\section{Proof of Eq.(\ref{ComTOk}) }
\label{Pro_ComTOk}
 For any $k\geq 1$,
according the following mappings for nodes of $\Lambda_k$ and $\Lambda_{k+1}$
\begin{eqnarray} \label{mapping}
\left \{                 
  \begin{array}{ll}
   A_{k+1}\equiv A_k, B_{k+1}\equiv O_k  & i_{k+1}=0 \\
   A_{k+1}\equiv O_k, B_{k+1}\equiv B_k & i_{k+1}=2m+1 \\
   A_{k+1}\equiv A_k, B_{k+1}\equiv C_k & i_{k+1}=1,2,\cdots,m \\
    A_{k+1}\equiv R_k, B_{k+1}\equiv B_k & i_{k+1}=m+1,m+2,\cdots,2m
  \end{array}
\right.
\end{eqnarray}
we have
\begin{eqnarray} \label{CaseofTB_A}
\hspace{-20mm}&&T_{B_{k+1}}-T_{A_{k+1}} \nonumber \\
\hspace{-20mm}&=&
\left \{                 
  \begin{array}{ll}
   T_{O_k}-T_{A_k} & i_{k+1}=0 \\
   T_{B_k}-T_{O_k} & i_{k+1}=2m+1 \\
   T_{C_k}-T_{A_k} & i_{k+1}=1,2,\cdots,m \\
   T_{B_k}-T_{R_k} & i_{k+1}=m+1,m+2,\cdots,2m
  \end{array}
\right.  \\
\end{eqnarray}
 Replacing $T_{O_k}$, $T_{C_k}$ and $T_{R_k}$ with Eqs.(\ref{TOk}), (\ref{TCk}) and (\ref{TRk}) respectively, we have
 \begin{eqnarray}
\hspace{-20mm}&&T_{B_{k+1}}-T_{A_{k+1}} \nonumber \\
\hspace{-20mm}&=& \label{HAK1_BK1}
\!\left \{  \!               
  \begin{array}{ll}
   \frac{1}{2}(T_{B_k}\!-\!T_{A_k})\!-\!2^{t-k}(2m\!+\!2)^{t-k-1}\!-\!2^{t-k-1}(2m\!+\!2)^{-k-1} & i_{k+1}=0 \\
   \frac{1}{2}(T_{B_k}-T_{A_k})\!+\!2^{t-k}(2m\!+\!2)^{t-k-1}\!+\!2^{t-k-1}(2m\!+\!2)^{-k-1}  & i_{k+1}=2m+1 \\
   \xi_k & i_{k+1}=1,2,\cdots,m \\
  -\xi_k & others 
  \end{array}
\right.
\end{eqnarray}
where $\xi_k=2^{t-k}[(2m\!+\!2)^t\!-\!(2m+2)^{t-k-1}]\!+\!2^{t\!-\!k\!-\!1}[1\!-\!(2m\!+\!2)^{-k-1}]$. \par
  For any $k\geq 1$, we find 
\begin{eqnarray}
  \hspace{-15mm}|T_{B_k}-T_{A_k}|\geq 2^{t-k+1}(2m+2)^{t-k}+2^{t-k}(2m+2)^{-k} \label{LTAK_BK} \\
  \hspace{-15mm}|T_{B_k}-T_{A_k}|\leq 2^{t-k+1}[(2m+2)^{t}-(2m+2)^{t-k}]+2^{t-k}[1-(2m+2)^{-k}]  \label{UTAK_BK}
\end{eqnarray}
The  Eqs.(\ref{LTAK_BK}) and (\ref{UTAK_BK}) are proved by mathematical induction as follows.\par
Let  $k=0$ in Eq. (\ref{HAK1_BK1}), we obtain
\begin{eqnarray}
\hspace{-20mm}|T_{B_{1}}\!-\!T_{A_{1}}|
& \!= \!&
\!\left \{  \!                
  \begin{array}{ll}
   \!2^{t}(2m\!+\!2)^{t-1}\!+\!2^{t-1}(2m\!+\!2)^{-1} & i_{1}=0 \\
   \!2^{t}(2m\!+\!2)^{t-1}\!+\!2^{t-1}(2m\!+\!2)^{-1} & i_{1}=2m+1 \\
   2^{t}[(2m\!+\!2)^t\!-\!(2m+2)^{t-1}]\!+\!2^{t\!-\!1}[1\!-\!(2m\!+\!2)^{-1}] &others
  \end{array}
\right.
\end{eqnarray}
Thus Eqs.(\ref{LTAK_BK}) and (\ref{UTAK_BK}) holds for $k=1$.\par
 Assuming that Eqs.(\ref{LTAK_BK}) and (\ref{UTAK_BK}) hold for some $k\geq 1$, we will prove   Eqs.(\ref{LTAK_BK}) and (\ref{UTAK_BK}) also hold for $k+1$. \par
 According to Eq.(\ref{CaseofTB_A}), $T_{B_{k+1}}-T_{A_{k+1}}$ has $2m+2$ cases due to the different value of $i_{k+1}$.
 It is easy to verify Eqs.(\ref{LTAK_BK}) and (\ref{UTAK_BK}) hold for  $i_{k+1}=0, 1,2,\cdots, 2m$ due to Eq.(\ref{HAK1_BK1}).\par
  For $i_{k+1}=0$, substituting  $T_{B_k}-T_{A_k}$ with right side of Eq.(\ref{LTAK_BK}), we obtain
\begin{eqnarray}
 \hspace{-20mm}|T_{B_{k+1}}-T_{A_{k+1}}|&= &|\frac{1}{2}(T_{B_k}-T_{A_k})\!-\!2^{t-k}(2m\!+\!2)^{t-k-1}\!-\!2^{t-k-1}(2m\!+\!2)^{-k-1}| \nonumber \\
  \hspace{-20mm}   &\geq & \frac{1}{2}\cdot \left[2^{t-k+1}(2m+2)^{t-k}+2^{t-k}(2m+2)^{-k}\right]\nonumber \\
  \hspace{-20mm}   &&  -2^{t-k}(2m\!+\!2)^{t-k-1}\!-\!2^{t-k-1}(2m\!+\!2)^{-k-1}\nonumber \\
  \hspace{-20mm}   &> &2^{t-k}(2m+2)^{t-k-1}+2^{t-k-1}(2m+2)^{-k-1} \label{LHAK1-BK1}
\end{eqnarray}
Substituting  $T_{A_k}-T_{B_k}$ with right side of Eq.(\ref{UTAK_BK})
, we have
\begin{eqnarray}
 \hspace{-20mm}|T_{B_{k+1}}-T_{A_{k+1}}|&= &|\frac{1}{2}(T_{B_k}-T_{A_k})\!-\!2^{t-k}(2m\!+\!2)^{t-k-1}\!-\!2^{t-k-1}(2m\!+\!2)^{-k-1}| \nonumber \\
  \hspace{-20mm}   &\leq & \frac{1}{2}\cdot \left[2^{t-k+1}[(2m+2)^{t}-(2m+2)^{t-k}]+2^{t-k}[1-(2m+2)^{-k}] \right]\nonumber \\
  \hspace{-20mm}   &&  +2^{t-k}(2m\!+\!2)^{t-k-1}\!+\!2^{t-k-1}(2m\!+\!2)^{-k-1}\nonumber \\
    \hspace{-20mm}   &= & 2^{t-k}[(2m+2)^{t}-(2m+2)^{t-k}+(2m\!+\!2)^{t-k-1}]\nonumber \\
  \hspace{-20mm}   &&  +2^{t-k-1}[1-(2m+2)^{-k}+(2m\!+\!2)^{-k-1}]\nonumber \\
  \hspace{-20mm}   &<&2^{t-k}[(2m\!+\!2)^t\!-\!(2m+2)^{t-k-1}]\!+\!2^{t\!-\!k\!-\!1}[1\!-\!(2m\!+\!2)^{-k-1}]\label{UHAK1-BK1}
\end{eqnarray}
Therefore, Eqs.(\ref{LTAK_BK}) and (\ref{UTAK_BK}) hold for $i_{k+1}=0$. \par
Similarity, we can prove they both hold for  $i_{k+1}= 2m+1$. Therefore, we obtain  Eqs.(\ref{LTAK_BK})  and (\ref{UTAK_BK})  hold for all the $2m+2$ cases of $T_{B_{k+1}}-T_{A_{k+1}}$  which led to they both hold for any $k\geq 1$.\par
We now come back to prove Eq.(\ref{ComTOk}). Without loss of generality,
 assuming $T_{B_k}\geq T_{A_k}$, according to Eq.(\ref{TOk}),  we obtain
   \begin{eqnarray}
 \hspace{-20mm}T_{O_{k}}-T_{A_k}&= &\frac{1}{2}(T_{B_k}-T_{A_k})\!-\!2^{t-k}(2m\!+\!2)^{t-k-1}\!-\!2^{t-k-1}(2m\!+\!2)^{-k-1}>0 
\end{eqnarray}
and
   \begin{eqnarray}
 \hspace{-20mm}T_{B_{k}}-T_{O_k}&= &\frac{1}{2}(T_{B_k}-T_{A_k})\!+\!2^{t-k}(2m\!+\!2)^{t-k-1}\!+\!2^{t-k-1}(2m\!+\!2)^{-k-1}>0 
\end{eqnarray}
Therefore,  Eq.(\ref{ComTOk}) holds while $T_{B_k}\geq T_{A_k}$. Similarity, we can prove  Eq.(\ref{ComTOk}) holds while $T_{B_k}\leq T_{A_k}$.
\section{Proof of Eq.(\ref{ComDOk}) }
\label{Pro_ComDOk}
 For any $k\geq 1$,
According the  mappings for nodes of $\Lambda_k$ and $\Lambda_{k+1}$ as shown in Eq.(\ref{mapping}), we have
\begin{eqnarray} \label{CaseofDB_A}
\hspace{-20mm}&&D_{B_{k+1}}-D_{A_{k+1}} \nonumber \\
\hspace{-20mm}&=&
\left \{                 
  \begin{array}{ll}
   D_{O_k}-D_{A_k} & i_{k+1}=0 \\
   D_{B_k}-D_{O_k} & i_{k+1}=2m+1 \\
   D_{C_k}-D_{A_k} & i_{k+1}=1,2,\cdots,m \\
   D_{B_k}-D_{R_k} & i_{k+1}=m+1,m+2,\cdots,2m
  \end{array}
\right.  \\
\end{eqnarray}
 Replacing $D_{O_k}$, $D_{C_k}$ and $D_{R_k}$ with Eqs.(\ref{DOk}), (\ref{DCk}) and (\ref{DRk}) respectively, we have
 \begin{eqnarray}
\hspace{-20mm}&&D_{B_{k+1}}-D_{A_{k+1}} \nonumber \\
\hspace{-20mm}&=& \label{RecDK}
\left \{                 
  \begin{array}{ll}
   \frac{1}{2}(D_{B_k}\!-\!D_{A_k})\!+\!2^{t-k-1}(2m+2)^{-k-1} & i_{k+1}=0\\
   \frac{1}{2}(D_{B_k}-D_{A_k})\!-\!2^{t-k-1}(2m+2)^{-k-1}  & i_{k+1}=2m+1 \\
   -2^{t-k-1}[1-(2m+2)^{-k-1}] &  i_{k+1}=1,2,\cdots,m \\
   2^{t-k-1}[1-(2m+2)^{-k-1}] & i_{k+1}=m+1,m+2,\cdots,2m
  \end{array}
\right.
\end{eqnarray}
  For any $k\geq 1$, we find
\begin{eqnarray}
  \hspace{-15mm}|D_{B_k}-D_{A_k}|\geq 2^{t-k}(2m+2)^{-k} \label{LDB_DAk} \\
  \hspace{-15mm}|D_{B_k}-D_{A_k}|\leq 2^{t-k}[1-(2m+2)^{-k}]  \label{UDB_DAk}
\end{eqnarray}
which are proved by mathematical induction as follows.\par
Let  $k=0$ in Eq. (\ref{RecDK}), we obtain
\begin{eqnarray}
\hspace{-10mm}D_{B_{1}}-D_{A_{1}}
&=&
\left \{                 
  \begin{array}{ll}
   2^{t-1}(2m\!+\!2)^{-1} & i_{1}=0 \\
   -2^{t-1}(2m\!+\!2)^{-1} & i_{1}=2m+1 \\
   2^{t\!-\!1}[1\!-\!(2m\!+\!2)^{-1}] & i_{1}=1,2,\cdots,m \\
   -2^{t\!-\!1}[1\!-\!(2m\!+\!2)^{-1}] & i_{1}=m+1,m+2,\cdots,2m
  \end{array}
\right.
\end{eqnarray}
It is easy to verify Eqs.(\ref{LDB_DAk}) and (\ref{UDB_DAk})  hold for $k=1$ .\par
 Assuming that (\ref{LDB_DAk}) and (\ref{UDB_DAk})  hold for some $k\geq 1$,  we will prove   Eqs.(\ref{LTAK_BK}) and (\ref{UTAK_BK}) also hold for $k+1$. \par
 According to Eq.(\ref{CaseofDB_A}), $D_{B_{k+1}}\!-\!D_{A_{k+1}}$ has $2m+2$ cases due to the different value of $i_{k+1}$.
 It is easy to verify Eqs.(\ref{LDB_DAk}) and (\ref{UDB_DAk})  hold for  $i_{k+1}=1,2,\cdots, 2m$ due to Eq.(\ref{RecDK}).\par
  For $i_{k+1}=0$, substituting  $D_{B_k}-D_{A_k}$ with right side of Eq.(\ref{UDB_DAk}) in  Eq.(\ref{RecDK}), we get
\begin{eqnarray}
\hspace{-20mm}|D_{B_{k+1}}-D_{A_{k+1}}| &=&|\frac{1}{2}(D_{B_k}-D_{A_k})\!+\!2^{t-k-1}(2m+2)^{-k-1}|\nonumber \\
   &\leq&\frac{1}{2}\cdot \left[2^{t-k}[1-(2m+2)^{-k}]\right]\!+\!2^{t-k-1}(2m+2)^{-k-1} \nonumber \\
   &< & 2^{t-k-1}[1-(2m+2)^{-k-1}]\label{UDAK1-BK1}
\end{eqnarray}
Substituting  $D_{A_k}-D_{B_k}$ with right side of Eq.(\ref{LDB_DAk}) in Eq.(\ref{RecDK}),  we have
\begin{eqnarray}
 |D_{B_{k+1}}-D_{A_{k+1}}|&=&|\frac{1}{2}(D_{B_k}-D_{A_k})\!+\!2^{t-k-1}(2m+2)^{-k-1}|\nonumber \\
   &\geq & \frac{1}{2}\cdot \left[2^{t-k}(2m+2)^{-k}\right]\!-\!2^{t-k-1}(2m+2)^{-k-1} \nonumber \\
   &> &2^{t-k-1}(2m+2)^{-k-1} \label{LDAK1-BK1}
\end{eqnarray}
Therefore, Eqs.(\ref{LDB_DAk}) and (\ref{UDB_DAk})  hold for $i_{k+1}=0$. Similarity, we can prove they both hold for  $i_{k+1}= 2m+1$. Therefore, we obtain  Eqs.(\ref{LDB_DAk}) and (\ref{UDB_DAk})   hold for all the $2m+2$ cases of $T_{B_{k+1}}-T_{A_{k+1}}$  which led to they both hold for any $k\geq 1$.\par
We now come back to prove Eq.(\ref{ComDOk}). Without loss of generality,
 assuming $D_{B_k}\geq D_{A_k}$, according to Eq.(\ref{DOk}),  we obtain
   \begin{eqnarray}
 \hspace{-20mm}D_{O_{k}}-D_{A_k}&=&\frac{1}{2}(D_{B_k}-D_{A_k})\!+\!2^{t-k-1}(2m+2)^{-k-1}>0 
\end{eqnarray}
and
   \begin{eqnarray}
 \hspace{-20mm}D_{B_{k}}-D_{O_k}&= &\frac{1}{2}(D_{B_k}-D_{A_k})\!-\!2^{t-k-1}(2m+2)^{-k-1} \nonumber \\
  \hspace{-20mm}   &\geq & \frac{1}{2}\cdot \left[2^{t-k}(2m+2)^{-k}\right]\!-\!2^{t-k-1}(2m+2)^{-k-1}\nonumber \\
 \hspace{-20mm}   &> &0
\end{eqnarray}
Therefore,  Eq.(\ref{ComDOk}) holds while $D_{B_k}\geq D_{A_k}$. Similarity, we can prove  Eq.(\ref{ComTOk}) holds while $D_{B_k}\leq D_{A_k}$.






\end{document}